\documentclass[11pt,a4paper]{article}

\usepackage[utf8]{inputenc}    
\usepackage[T1]{fontenc}       
\usepackage{lmodern}           
\usepackage{amsmath,amssymb}   
\usepackage{graphicx}          
\usepackage{hyperref}          
\usepackage{authblk}           
\usepackage{geometry}          
\usepackage{multirow} 

\usepackage[numbers,longnamesfirst]{natbib}
\usepackage[draft,inline,marginclue]{fixme}
\usepackage[utf8]{inputenc}
\usepackage[htt]{hyphenat}

\usepackage{listings}
\usepackage{xcolor}

\definecolor{backcolour}{rgb}{0.95,0.95,0.92}

\lstdefinestyle{mystyle}{
    commentstyle=\color{blue},
    keywordstyle=\color{red},
    numberstyle=\tiny\color{black}, 
    stringstyle=\color{magenta},
    basicstyle=\ttfamily\footnotesize,
    breakatwhitespace=false,         
    breaklines=true,                 
    captionpos=b,                    
    keepspaces=true,                 
    numbers=left,                     
    numbersep=5pt,                    
    showspaces=false,                
    showstringspaces=false,
    showtabs=false,                  
    tabsize=2,
    stepnumber=1,                     
    frame=single                      
}

\lstdefinestyle{openfoamStyle}{
    language=C++,
    basicstyle=\ttfamily\footnotesize,
    keywordstyle=\color{magenta},
    commentstyle=\color{teal},
    stringstyle=\color{orange},
    numbersep=5pt,
    breaklines=true,
    captionpos=b,
    frame=single,                      
    tabsize=4,
    morekeywords={Info, endl, simple, runTime, fvScalarMatrix, fvm, fvOptions, TEqn}
}

\lstdefinestyle{openfoamStyle_double}{
    float,
    language=C++,
    basicstyle=\scriptsize,
    keywordstyle=\color{magenta},
    commentstyle=\color{teal},
    stringstyle=\color{orange},
    numbersep=5pt,
    breaklines=true,
    captionpos=b,
    frame=single,                      
    tabsize=4,
    morekeywords={Info, endl, simple, runTime, fvScalarMatrix, fvm, fvOptions, TEqn}
}

\usepackage{subcaption}
\newcommand{\meshs}{\texttt{Mesh-S} }
\newcommand{\meshm}{\texttt{Mesh-M} }
\newcommand{\meshl}{\texttt{Mesh-L} }
\newcommand{\meshxl}{\texttt{Mesh-XL} }

\newcommand{\cascadelake}{\texttt{CascadeLake} }
\newcommand{\epito}{\texttt{EpiTo} }
\newcommand{\gracehopper}{\texttt{Grace-Hopper} }
\newcommand{\twophases}{\texttt{Two-Phases} }

\newcommand{\laplacianfoam}{\texttt{laplacianFoam} }

\begin{document}
\listoffixmes
\title{Building an Accelerated OpenFOAM Proof-of-Concept Application using Modern C++}

\author[1]{Giulio Malenza\thanks{Corresponding author: \href{mailto:giulio.malenza@unito.it}{giulio.malenza@unito.it}. ORCID: \href{https://orcid.org/0009-0006-4862-7429}{0009-0006-4862-7429}}}
\author[2]{Giovanni Stabile\thanks{\href{mailto:giovanni.stabile@santannapisa.it}{giovanni.stabile@santannapisa.it}. ORCID: \href{https://orcid.org/0000-0003-3434-8446}{0000-0003-3434-8446}}}
\author[3]{Filippo Spiga\thanks{\href{mailto:fspiga@nvidia.com}{fspiga@nvidia.com}. ORCID: \href{https://orcid.org/0000-0003-1448-5304}{0000-0003-1448-5304}}}
\author[1]{Robert Birke\thanks{\href{mailto:robert.birke@unito.it}{robert.birke@unito.it}. ORCID: \href{https://orcid.org/0000-0003-1144-3707}{0000-0003-1144-3707}}}
\author[1]{Marco Aldinucci\thanks{\href{mailto:marco.aldinucci@unito.it}{marco.aldinucci@unito.it}. ORCID: \href{https://orcid.org/0000-0001-8788-0829}{0000-0001-8788-0829}}}

\affil[1]{Department of Computer Science, University of Torino, Corso Svizzera 185, 10149 Torino, Italy}
\affil[2]{Biorobotics Institute, Sant'Anna School of Advanced Studies, Viale Rinaldo Piaggio 34, 56025 Pisa, Italy}
\affil[3]{NVIDIA Corporation, 2788 San Tomas Expressway, Santa Clara, CA 95051, United States}

\maketitle
\begin{abstract}
The modern trend in High-Performance Computing (HPC) involves the use of accelerators such as Graphics Processing Units (GPUs) alongside Central Processing Units (CPUs) to speed up numerical operations in various applications. Leading manufacturers such as NVIDIA, Intel, and AMD are constantly advancing these architectures, augmenting them with features such as mixed precision, enhanced memory hierarchies, and specialised accelerator silicon blocks (e.g., Tensor Cores on GPU or AMX/SME engines on CPU) to enhance compute performance. At the same time, significant efforts in software development are aimed at optimizing the use of these innovations, seeking to improve usability and accessibility. This work contributes to the state-of-the-art of OpenFOAM development by presenting a working Proof-Of-Concept application built using modern ISO C++ parallel constructs.
This approach, combined with an
appropriate compiler runtime stack, like the one provided by the NVIDIA HPC SDK, 
makes it possible to accelerate well-defined kernels, allowing multi-core execution and GPU offloading using a single codebase. The study demonstrates that it is possible to increase the performance of the OpenFOAM laplacianFoam application by offloading the computations on NVIDIA GPUs using the C++ parallel construct.
\end{abstract}

\vspace{1em}
\noindent\textbf{Keywords:} OpenFOAM; High-performance computing; GPU; Multicore; ISO C++ language; Standard library.

\section{Introduction}

The current trend in High-Performance Computing (HPC) includes the use of accelerators, such as Graphics Processing Units (GPUs), as co-processors coupled with Central Processing Units (CPUs) to accelerate numerical simulations in many fields of science and engineering \cite{GPUComputing1}. From a hardware point of view, technology providers such as NVIDIA, Intel, and AMD are focused on developing and improving these new technologies, adding new functionalities such as mixed precision, better memory hierarchies, and tensor core for high throughput in each generation. From the software point of view, many efforts have been made to make the best use of these new technologies and make them easy for users to handle. Despite all these efforts, porting existing codebases to GPU architectures remains a complex and often non-trivial task \cite{Aldinucci_Practical_P_2021,SALVADORE2013129}. Although vendors provide a variety of libraries, domain-specific frameworks, and parallel programming models to make it easier for developers to perform this task, leveraging them frequently necessitates the rewriting or adaptation of significant portions of code. This, in turn, can compromise the portability of the application on different hardware platforms and hinder long-term maintenance \cite{Pennycook,REGULY2020104425}. 

One of the most widely used open-source software packages for Continuum Mechanics and Computational Fluid Dynamics (CFD) is OpenFOAM\footnote{\url{https://www.openfoam.com/}} \cite{OrigianlOF,JASAK2}. It is widely used in both academia and industry for simulating a wide range of multi-physics phenomena. One common aspect of all is being computationally intensive \cite{Ouro_2021} as well as memory bandwidth sensitive, making GPU a very attractive platform for obtaining a noticeable improvement in time to solution.

OpenFOAM is written in object-oriented C++ and mainly parallelized with the Message Passing Interface (MPI). Its open-source nature allows users to access, modify, and extend the codebase, making it highly customizable for a broad range of engineering applications. However, its extensive use of macros and template classes brings both advantages (e.g. extendability, class composability, readable high-level syntax) and few drawbacks. Due to the large codebase (on the order of one million lines) and the extreme reliance on C++ abstract constructs, porting OpenFOAM to GPUs presents a considerable challenge \cite{tandon2024porting}, which requires balancing both performance and portability.

In fact, a key strength of OpenFOAM is its portability across CPU architectures, largely thanks to the portable nature of C++ (a language supported by many compilers across many computing architectures). The main challenge, therefore, lies in enabling GPU support without sacrificing portability or introducing excessive code duplication and specialization for different hardware vendors. The proposed solution is transition to a codebase written in compliant future-proof modern C++ using C++ parallel algorithms without requiring massive intrusive code changes or use of non-standard extensions.

Starting with C++11 \cite{cpp6}, the language introduced its first set of features for parallelism and concurrency. Subsequent revisions of the standard - C++14, C++17, and C++20 - further consolidate and expand these capabilities. Notably, C++17 was the first iteration to officially extended the Standard Template Library (STL) with parallel algorithms and execution policies, enabling developers to express and exploit parallel computation more effectively by just writing C++ code\cite{cpp7}.

\subsection{Article Overview} 

This work explores the use of modern ISO C++ parallel constructs to accelerate key routines and functionalities. To demonstrate the feasibility of this approach, we built a simple Proof of Concept inspired by \laplacianfoam, one of the simplest OpenFOAM applications. Key core routines and classes have been refactored to leverage C++17 and newer constructs, as there is no \textit{free lunch} in GPU porting (in any way or another, some code needs to evolve to expose parallelism). The novelty of this work consists of proving that adopting the C++ Parallel Standard Template Library (PSTL) unlocks both multicore and GPU execution with just a compiler flag switch without requiring domain experts to have any special technical knowledge.

The following section provides background information and a review of related work. Subsequently, a concise overview of the ISO C++ Standard and the parallel programming techniques developed in recent years is presented. Sections four and five delve into the most relevant features of OpenFOAM for this study and the implementation methodology used. The experimental results and concluding remarks are discussed in sections six and seven.
\section{Releated Works}

\subsection{Parallel Programming Models}
With the emergence of heterogeneous architectures in the past twenty years, many parallel programming frameworks have been developed to exploit parallel features of the new technologies. Such frameworks can be subdivided into three categories. The low-level parallel programming frameworks, such as CUDA~\cite{Kirk_2007}, HIP~\cite{Kerscher_2022}, and SYCL~\cite{Johnston_2020}, are language-specific models exposing dedicated APIs for fine-tuning, pre-fetching, and memory data transfers. Typically, these frameworks generate long and complex code, which is not always portable and easy to maintain. Pragma-based frameworks, like OpenMP~\cite{Huber_2022} and OpenACC~\cite{Herdman_2014}, are based on \texttt{\#pragma} directives that instruct the compiler to generate assembly code running on multi-core or GPU systems. These frameworks usually guarantee higher portability across different platforms and easier maintenance. However, they could become verbose and difficult to read \cite{Aldinucci_Practical_P_2021,OpenMPSalvadore}. Moreover, they have limited APIs for memory management and kernel fine-tuning. The third category of parallel programming models consists of C++ abstraction libraries. These frameworks have no directives to specify data transfers between CPU and GPU, and no possibility of fine-tuning kernels. However, they allow users to write code that is extremely portable across different platforms and easy to maintain. Examples of such libraries are KOKKOS~\cite{Edwards_2014}, RAJA~\cite{Beckingsale_2019}, Alpaka~\cite{Zenker_2016}, Thrust~\cite{THRUSTBook}, 
OCCA~\cite{Medina_2014},
FastFlow~\cite{Aldinucci_2010}, SkePU~\cite{SkePU} and C++ PSTL~\cite{DeLozier_2012}. This work focuses on C++ PSTL, an open standard that relies only on the standard library, fully abstracting any underlying parallel runtime. 

\subsection{Past OpenFOAM Porting Efforts}

The historical trajectory of GPU acceleration in OpenFOAM reveals a pattern of pioneering, yet ultimately fragmented, efforts only partially address performance gaps while exposing unresolved challenges. Early initiatives focused predominantly on linear solver offloading~\cite{CulpoMas, A01, A02} while neglecting critical computational phases. The \textit{PETSc4FOAM}~\cite{petsc4foam}, a plug-in library to include PETSc~\cite{petscGropp} into the OpenFOAM framework, allows efficient parallel execution of the solver. This plug-in can also leverage NVIDIA \texttt{AmgX} library~\cite{AMGX-lib} to offload the bulk of the computation on GPUs~\cite{amgx}. The \texttt{AmgX} library currently supports only NVIDIA GPUs. Recent benchmarks demonstrate $8$x–$10$x speedups for linear solvers, though overall simulation gains remain constrained to $1.7$x–$2.2$x due to unaddressed bottlenecks in matrix construction (22–25\% runtime) and I/O operations (12–15\%). Industrial validations on Leonardo's GPU clusters show $4.5$x acceleration for aerodynamic simulations, yet highlight persistent PCIe transfer overheads consuming 15–20\% of multi-GPU runtime.

Beyond the focus on the solver, multiple porting efforts have been produced, each one with its unique set of successes. The SpeedIT plugin by Wroclaw University researchers achieved full GPU-accelerated SIMPLE/PISO solvers for OpenFOAM 2.x, demonstrating $4.2x$ speedups over contemporary CPUs through optimized sparse matrix formats and batched GPU operations~\cite{tomczak2012complete}. Lukarski's PARALUTION plugin~\cite{lukarski2013paralution} enabled multi-GPU sparse iterative solvers through intrusive OpenFOAM modifications, though synchronization overheads and fixed precision constraints reduced practical utility. Industrial implementations such as FluiDyna's Culises and simFlow's RapidCFD~\cite{RapidCFD} achieved $2.6$x–$3.5$x end-to-end speedups via GPU-accelerated matrix assembly, but the lack of an upstreaming strategy and long-term maintenance resulted in abandoned codebases still publicly available but of no longer practical use. \texttt{RapidCFD} is an open-source OpenFOAM implementation that runs almost all simulations on NVIDIA GPUs. It is based on an outdated version of OpenFOAM. RapidCFD uses the \texttt{Thrust} library~\cite{NVIDIA7} to accelerate the assembly step. Thrust is a robust library of parallel algorithms and data structures written in C++ with a CUDA back-end. Thrust is part of the NVIDIA CUDA Core Compute Libraries (CCCL) suite. 

In recent years, more forward-looking solutions have focused on hybrid GPU-CPU workflows and unified memory architectures. In addition to the CUDA programming model, other programming models have also been explored. AMD developed a proof-of-concept version of OpenFOAM OpenMP offload~\cite{tandon2024porting} that targets the AMD GPU MI300A product. The \texttt{zeptoFOAM}\footnote{\url{https://gitlab.hpc.cineca.it/exafoam/zeptoFOAM}} and \texttt{NeoN}\footnote{\url{https://github.com/exasim-project/NeoN}} are two recently funded initiatives aimed at re-implementing several core OpenFOAM classes for better acceleration support, introducing device-aware memory management and GPU-optimized routines. \texttt{zeptoFOAM} adopts CUDA and \texttt{AmgX} (for the linear solver) while \texttt{NeoN} uses KOKKOS~\cite{Edwards_2014} and Ginkgo~\cite{ginkgo-toms-2022} (for the linear solver).

The foundation of our work is based on portability, compliance to the C++ standard, and development speed being as important criteria as pure performance in order to guarantee longevity and maintainability of any new code contribution. The adoption of ISO C++ standard parallelism is an ideal fit.
\section{ISO Standard C++}\label{sec:cpp}

Multi-threading in C++ was officially introduced in ISO C++11. This standard also introduced the memory consistency model, which describes the allowed behavior of multi-threaded programs running with shared memory~\cite{cpp1}. Multi-threading features have been further expanded in the ISO C++14, C++17, and C++20 standards~\cite{cpp2,cpp3}. Particularly important for this work is the ISO C++17 standard, which introduced the Parallel Standard Template Library (PSTL).

Initially, the PSTL~\cite{cpp4,cpp5} comprised three components: containers, consisting of objects that store a collection of other objects; algorithms that run on the containers; and iterators, which allow algorithms to run over the container elements. In the Standard C++11, the STL counted about 80 algorithms. At present, the count is around 100. These can be classified into different categories according to their features. A class of algorithms allow one to iterate over or transforming container elements. In this class, a popular key algorithm used in this work is the \texttt{for\_each} algorithm. This algorithm applies a function pointer to each element in the container over which it iterates. Another essential algorithm in this class is the \texttt{transform} algorithm, which applies a given operation to a range and returns the result in another range. Another essential class of algorithms can be used to perform summary operations. This class includes algorithms such as \texttt{ count}, which counts the number of elements that satisfy some condition in a given range, and \texttt{reduce}, which applies a reduction operation on a range of elements. Other algorithms like \texttt{search} and \texttt{find} allow identifying elements in various containers. Another class of algorithms allows one to manipulate the container's memory and initialization. Among them, three relevant algorithms are: \texttt{copy}, which copies all elements in a range into another range, \texttt{generate}, which assigns to each element in a range a value generated by a given function object, and \texttt{fill} which assigns a specific value to each element in a range. Finally, algorithms like \texttt{sort, stable\_sort, partial\_sort} allow sorting elements in a range. 

When the standard C++17 was introduced, many STL algorithms were extended to include the possibility to specify the \texttt{execution policy} when invoking them. By selecting an execution policy, the target algorithm can be executed in multiple ways: sequentially, sequentially with vectorization (from C++20 onward), parallel, and parallel with vectorization. Specifically:
\begin{itemize}
    \item \texttt{std::execution::seq} $\rightarrow$ from C++17. It forces the execution of an algorithm to run sequentially on the CPU.
    \item \texttt{std::execution::unseq} $\rightarrow$ from C++20. The calling algorithm is executed using vectorization on the calling thread.
    \item \texttt{std::execution::par} $\rightarrow$ from C++17. This policy suggests to the compiler that the algorithm can be executed in parallel.
    \item \texttt{std::execution::par\_unseq} $\rightarrow$ from C++17. It suggests that the algorithm can be executed in parallel on multiple threads, each capable of vectorizing the calculus.
\end{itemize}

It is worth noting that parallel execution policies allow the system to perform the calculation with multiple threads, but this is not a requirement. The standard C++ leaves the compiler with great freedom in deciding if, when, and how to run the algorithms in parallel.

Another important point is that parallel algorithms do not automatically protect the user from data races and deadlocks. It is the user's responsibility to ensure that changing the execution policy, typically from sequential to parallel, does not introduce deadlocks or data races.

Different compilers can compile and execute C++ algorithms in parallel, specifying the correct execution policy.
The NVIDIA HPC SDK is a comprehensive toolbox for GPU acceleration of modeling and simulation applications. It includes compilers for C, C++, and Fortran, numerical libraries, and analysis tools necessary for developing HPC applications on the NVIDIA platform and allows the mixing of various programming languages.
Unlike GNU, \texttt{nvc++} can compile and generate an executable able to run in multi-threading on CPUs and GPUs via a simple compilation flag.  Adding \texttt{-stdpar=gpu} generates kernels for GPU executions, while specifying \texttt{-stdpar=multicore} generates code executed via multi-threading on the CPU. 
Recent studies~\cite{DeakinNBodySim,nvidia1,nvidia2} showed that using this new parallel programming approach can lead to cleaner code and performance gains. An example of this is \textit{Lulesh}, a hydrodynamics mini-app from Lawrence Livermore National Laboratory (LLNL) written in C++ and then modified to use standard parallelism with NVC++~\cite{nvidia3}. Performance obtained using standard C++ parallelism with offload on to NVIDIA A100 GPU is around $13x$ faster than using OpenMP offload on an AMD EPYC 7742. 

One of the key points in developing code for GPUs is how data movement between CPU and GPU takes place. CUDA and OpenACC have specific directives that allow the user to manage this data movement. The ISO C++ does not explicitly include any construct or operation to manage data movement between two distinct computing devices. Allowing so would break the C++ memory model and the C++ abstract machine. In order to allow programming GPU using standard C++ parallel constructs, the data movement management has to be entirely transparent and self-managed. To do so, \texttt{nvc++} assumes CUDA Unified Memory~\cite{nvidia4} is enabled by default.

Unified memory creates a single Unified Virtual Address space accessible from any processor in a system~\cite{nvidia5,HarnessingGH200}. This allows the allocation of a pool of managed memory that is shared between CPU and GPU. Data allocated with this approach can be read and written from code running on CPUs or GPUs. The NVIDIA CUDA driver and NVIDIA GPU hardware are responsible, via a page-fault mechanism, for  triggering data migration. 
Only data dynamically allocated on the CPU can be marked and managed. Hence, data allocated on the CPU or GPU stack cannot be automatically moved. Any de-referenced pointer and referenced object in a parallel C++ algorithm must refer to the CPU heap.
\section{OpenFOAM}
OpenFOAM~\cite{JASAK200989,OFuserguide} is one of the most popular open-source Computational Fluid Dynamics (CFD) software due to its flexibility and other optimized features such as extensible solvers, efficient parallel computing, dynamic mesh handling, and advanced optimization features for customizable and high-performance simulations. It is mainly written in C++ and is based on the finite-volume method (FVM)~\cite{FVM,OpenFOAM_BIBLE,ThesisOF}. The official release is parallelised using a message-passing programming model. Any
OpenFOAM simulation consists of two main parts: the assembly part, where linear systems of algebraic equations are derived from the discretization of a set of Partial Differential Equations (PDEs), and the solver part, in which the solution of these systems is obtained using numerical methods. Discretizing and solving these equations is difficult not only from a theoretical but also a computational point of view.

\subsection{The \emph{Assembly} phase}
OpenFOAM uses FVM to represent and evaluate general PDEs through algebraic equations. Among the most important PDEs of fluid dynamics are the Navier-Stokes equations. To solve these equations, OpenFOAM first discretises them as a set of algebraic equations and then solves them using a numerical method. The discretisation phase is traditionally called the assembly phase. This work focuses on this phase and, in particular, on the assembly operators used in the \verb|laplacianFoam| application. More details can be found in the books~\cite{OpenFOAM_BIBLE,FVM} and in the Ph.D. thesis~\cite{ThesisOF}.

The numerical solution of a partial differential equation consists of finding the values of a dependent variable at specified points called grid elements. Therefore, the main goal is to replace the continuous exact solution of the equation considered with discrete values at those points. Typically, the result of the geometric discretisation is a mesh. This mesh consists of a discrete set of non-overlapping elements that fill the entire physical domain. There can be structured and unstructured meshes. The FVM implemented in OpenFOAM allows using mesh elements of arbitrary convex polyhedral shapes. 
Elements can also be defined in terms of their bounded faces. Usually, the faces are subdivided into interior faces, which connect two elements, and boundary faces, which coincide with the domain's boundary.

\begin{figure}[h!]
    \centering
    \includegraphics[scale=0.27]{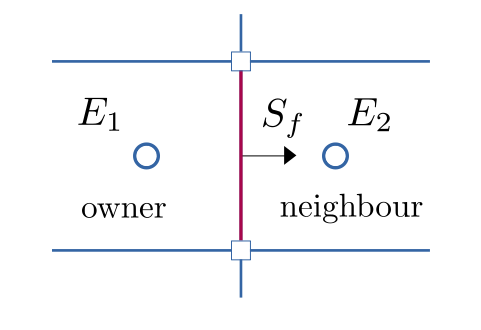} 
    \caption{Representation of owners and neighbours. }
    \label{fig:OwnerNeighbour}
\end{figure}

The assembly phase can be subdivided into two main parts: local and global assembly. During the local part, the equations are integrated over each element. After this, the system of linear equations can be built in the global assembly using the previous contributions. In the case of structured meshes, local indices are automatically mapped into global indices. This simplification is due to the geometric symmetries of the domain under consideration.
The situation is more complex for unstructured meshes. Since the elements can be of arbitrary polyhedral shape, defining indices that map local contributions into global ones is not possible. Topological information about elements, faces, and vertices is needed to compute the assembly. This information is represented in terms of connectivity lists. Specifically, there are element, face, and vertex connectivities. OpenFOAM addresses this problem using a face-addressing storage method, which uses a set of lists (array) to store points, faces, and elements. Following the dictates of this method, OpenFOAM usually stores all relevant information about the mesh in the folder \texttt{constant/polyMesh}. This folder contains the following files \cite{OpenFOAM_UserGuide}:
\begin{itemize}
    \item \texttt{points}: the list of vectors describing the coordinates of the elements vertices.
    \item \texttt{faces}: the list of faces, each face is identified by a list of indices to vertices in the points list.
    \item \texttt{owner}: a list of owner element labels, the index of entry relating directly to the face.
    \item \texttt{neighbour}: a list of neighbour element labels.
    \item \texttt{boundary}: a list of patches.
\end{itemize}
\texttt{Owner} and \texttt{neighbour} lists are significant for this work. The \texttt{owner} is a list of sizes and the number of total faces. Each entry of this list represents the owner element label of that face. Figure~\ref{fig:OwnerNeighbour} represents the face connectivity. The element \texttt{$E_1$} represents the \textit{owner} of the face \texttt{$f$} while the element \texttt{$E_2$} is its neighbour. The surface normal vector points from the owner to the neighbour elements.
The neighbour is a list where each entry represents the element label neighbour of that face. It is important to note that each face has an owner element label, but not every face has a neighbour element label. Indeed, boundary faces have no neighbours. For this reason, the size of the neighbour list is equal to the number of neighbour faces.

\section{Proof-of-Concept Implementation}
\subsection{The \laplacianfoam application}\label{laplacianFOAM}

One of the most basic and simple OpenFOAM applications is \laplacianfoam. The application is used primarily for solving scalar transport equations of a scalar field. It is represented by equation~\ref{eq:lap}, where $T$ is the scalar field (the temperature in our case).
\begin{equation}
    \frac{\partial}{\partial T}(T) -\nabla\cdot(D_T\nabla T)=S_T.
    \label{eq:lap}
\end{equation}
The first term is the time derivative of $T$ and represents the transient behavior; the second term is the diffusion operator representing the diffusion of T. It comprises the diffusion coefficient $D_T$ and the temperature gradient $\nabla T$. The term on the right is the source term $S_T$.
Listing~\ref{lst:laplacianFoam} shows how the equation is written and solved in OpenFOAM. In line 4, the \texttt{SIMPLE} loop is used to iterate until the residual is small. Inside that loop, another loop helps approximate numerical solutions in the case of non-orthogonal meshes. In this loop, the gradient computation is divided into an implicit contribution and an explicit contribution; more details on this correction are available in \cite{ThesisOF}. Lines 10-15 represent the assembly phase of the equation. Specifically, on line 12, \texttt{fvm::ddt(T)} is the time derivative of the scalar field T and \texttt{fvm::laplacian(DT, T)} represents the Laplacian term where $D_T$ is the diffusion coefficient. Line 14 defines an additional source term. Line 17 adjusts the equation, considering constraints if defined, and line 18 solves the equation. Line 19 corrects the solution if other options are specified. Line 22 and 24 write the solution on file and print the execution time information.

\vbox{\lstset{style=openfoamStyle}
\begin{lstlisting}[language=C++,caption={\laplacianfoam application}, label={lst:laplacianFoam}]
Info<< "\nCalculating temperature distribution\n" << endl;

while (simple.loop())
{
    Info<< "Time = " << runTime.timeName() << nl << endl;

    while (simple.correctNonOrthogonal())
    {
        fvScalarMatrix TEqn
        (
            fvm::ddt(T) - fvm::laplacian(DT, T)
         ==
            fvOptions(T)
        );

        fvOptions.constrain(TEqn);
        TEqn.solve();
        fvOptions.correct(T);
    }

    #include "write.H"

    runTime.printExecutionTime(Info);
}

Info<< "End\n" <<endl;
\end{lstlisting}}

In order to apply the techniques explained in Section \ref{sec:cpp}, we looked at the most computationally intensive functions outside the solver. We used \texttt{Valgrind}~\cite{Valgrind} to precisely identify which routines to prioritize for the porting effort.

The total execution time of the application can be divided into four main components. The portion associated with the \texttt{fvMesh} function accounts for $13.84$\% of the total time and corresponds to the generation of the mesh and the initialization of the field. The write phase, which represents 8.97\% of the time, is dedicated to storing the computed solution. The \texttt{solve} phase consumes 31.47\% of the total time and encompasses the solution of the linear system. Finally, the \texttt{fvm::Laplacian} component, which represents the evaluation of the Laplacian operator, accounts for the largest share at 40.31\%.

Mesh creation and data writing can become performance bottlenecks in complex simulations. However, this study primarily examines the routines and operators involved in the assembly step, which is the main bottleneck when running simulations on GPUs. When only the solver phase is offloaded to the GPU and the assembly phase remains on the CPU, the necessary data transfers between the CPU and GPU introduce significant overhead. This overhead can severely reduce performance, effectively negating the advantages gained by executing the solver phase on the GPU.

While the focus of this work was on identifying and demonstrating a strategy to port the assembly stage, we have also accelerated the native OpenFOAM Preconditioner Conjugate Gradient (PCG). By doing so, we could demonstrate a complete end-to-end solution running on GPU.

\subsection{Code Porting}

The source code is available at \url{https://github.com/alpha-unito/OpenFOAM}. The \texttt{main} branch contains the reference OpenFOAM code (version 24.12) and the \texttt{of-stdpar} contains all changes. A \texttt{README\_STDPAR.md} explains how to compile, run, and tune the execution of the \texttt{laplacianFoam} test.

Due to space limitations, a comprehensive summary of all code modifications cannot be included. Instead, the following section highlights key changes introduced to offload most of the gradient computation. The techniques presented are indicative of the broader set of transformations applied to the other operators examined in this study.

The gradient operator is evaluated through the \texttt{calcGrad} function that calls three other major functions (and respected classes):
\begin{itemize}
\item \texttt{\textbf{interpolate}} (class: \texttt{surfaceInterpolationScheme}).
\item \texttt{\textbf{gradf}} (class:) \texttt{gaussGrad}).
\item \texttt{\textbf{correctBoundaryConditions}} (class: \texttt{GaussGrad}).
\end{itemize}


\subsubsection{Class: \texttt{SurfaceInterpolationScheme}}
\noindent The \texttt{SurfaceInterpolationScheme} is a class implemented in the finite-volume library that performs interpolation from volume fields to face fields. The member function \texttt{dotInterpolate} was modified.
Its main loop, see Listing \ref{lst:dotInterpolate}, runs sequentially on the CPU. 

\vspace{0.2cm}
\vbox{\lstset{style=openfoamStyle}
\begin{lstlisting}[language=C++,caption={\texttt{dotInterpolate} loop}, label={lst:dotInterpolate}]
for (label fi=0; fi<P.size(); fi++)
{
  sfi[fi] = Sfi[fi] & (lambda[fi]*(vfi[P[fi]] 
                    - vfi[N[fi]]) + vfi[N[fi]]);
}
\end{lstlisting}}

Using the \texttt{for\_each} algorithm, iterating in parallel over an iterator generated by \texttt{iota} and performing the calculation within a \texttt{lambda}, the loop can be performed in parallel (see Listing \ref{lst:dotInterpolatepar}).

\vspace{0.2cm}
\vbox{\lstset{style=openfoamStyle}
\begin{lstlisting}[language=C++,caption={\texttt{dotInterpolate} parallel loop}, label={lst:dotInterpolatepar}]
auto* Sfii = &Sfi; 
auto iter=std::views::iota(0,P.size());

std::for_each(std::execution::par,iter.begin(),iter.end(),
    [N,P,l=lambda.cdata(),Sfii,v=vfi.cdata(),s=sfi.data()](const auto& fi){
      s[fi] = (*Sfii)[fi] & (l[fi]*(v[P[fi]] 
                            - v[N[fi]]) + v[N[fi]]);
    }
);
\end{lstlisting}}

Another member function in this phase is the \texttt{weights}. This function computes the weights required for the interpolation by the previous \texttt{dotInterpolate}. This is only called the first time during a general simulation unless the mesh changes. Its main loop was parallelized using the \texttt{transform} algorithm as follows:

\vspace{0.2cm}
\vbox{\lstset{style=openfoamStyle}
\begin{lstlisting}[language=C++,caption={\texttt{weights} parallel loop}, label={lst:weightspar}]
auto iter=std::views::iota(0,owner.size());

std::transform( std::execution::par, iter.begin(), iter.end(), w.begin(),[ ne=neighbour.cdata(), ow=owner.cdata(), s=Sf.cdata(), c=C.cdata(), cf=Cf.cdata()] (const auto& facei){
    scalar SfdOwn = mag(s[facei] & (cf[facei] - c[ow[facei]]));
    scalar SfdNei = mag(s[facei] & (c[ne[facei]]- cf[facei]));
    if (mag(SfdOwn + SfdNei) > ROOTVSMALL){
      return SfdNei/(SfdOwn + SfdNei);
    } else {
      return 0.5;
    } 
  }
);
\end{lstlisting}}

\subsubsection{Function: \texttt{gradf}}
The second and most computationally intensive phase during gradient evaluation is the \texttt{gradf} member function, part of the \texttt{gaussGrad} class.
At the beginning of the computation, a new temporary field is constructed, see Listing~\ref{lst:buildGrad}. This constructor calls other functions and, at last, invokes the field constructor of the class \texttt{Field}. This constructor is responsible to allocates the necessary CPU memory and inizialize the field with a sequential loop on CPU.

\vspace{0.2cm}
\vbox{\lstset{style=openfoamStyle}
\begin{lstlisting}[language=C++,caption={Gradient field constructor}, label={lst:buildGrad}]
tmp<GradFieldType> tgGrad(
  new GradFieldType(
   IOobject(
     name,
     ssf.instance(),
     mesh,
     IOobject::NO_READ,
     IOobject::NO_WRITE
   ),
   mesh,
   dimensioned<GradType>(ssf.dimensions()/dimLength,Zero),
   extrapolatedCalculatedFvPatchField<GradType>::typeName
  )
);
\end{lstlisting}}

This loop can be translated and executed in parallel using a \texttt{fill\_n} algorithm, see Listing~\ref{lst:fieldBuildPar}.

\vspace{0.2cm}
\vbox{\lstset{style=openfoamStyle}
\begin{lstlisting}[language=C++,caption={\texttt{Field} constructor in parallel.}, label={lst:fieldBuildPar}]
template<class Type>
inline Foam::Field<Type>::Field(const label len, const Type& val):List<Type>(len)
{
  std::fill_n(std::execution::par,this->begin(),len,val);
}
\end{lstlisting}}

The second intensive computation in the \texttt{gradf} function is the loop shown in Listing~\ref{lst:firstcycle}.

\vspace{0.2cm}
\vbox{\lstset{style=openfoamStyle}
\begin{lstlisting}[language=C++,caption={First cycle in the \texttt{gradf} routine.}, label={lst:firstcycle}]
forAll(owner, facei)
{
    const GradType Sfssf = Sf[facei]*issf[facei];
    igGrad[owner[facei]] += Sfssf;
    igGrad[neighbour[facei]] -= Sfssf;
}
\end{lstlisting}}

This loop cannot be trivially parallelised because parallel execution of independent threads performing different iteration count may lead to a data race condition. The most immediate way to solve this inconvenience would be to use atomic operations for critical value updates.
The performance of atomic operations can vary greatly depending on how often two threads have to write to the same memory address. If this frequency is low, good performance is achieved; in the worst case, this leads to sequential execution.

We propose and implement a re-organisation of the \texttt{owner} and \texttt{neighbour} lists to make them "atomic free". We demonstrate the idea with the \texttt{owner} list, the same procedure is applied to the \texttt{neighbour} list. 

\begin{figure}[h!]
    \centering
    \includegraphics[scale=0.45]{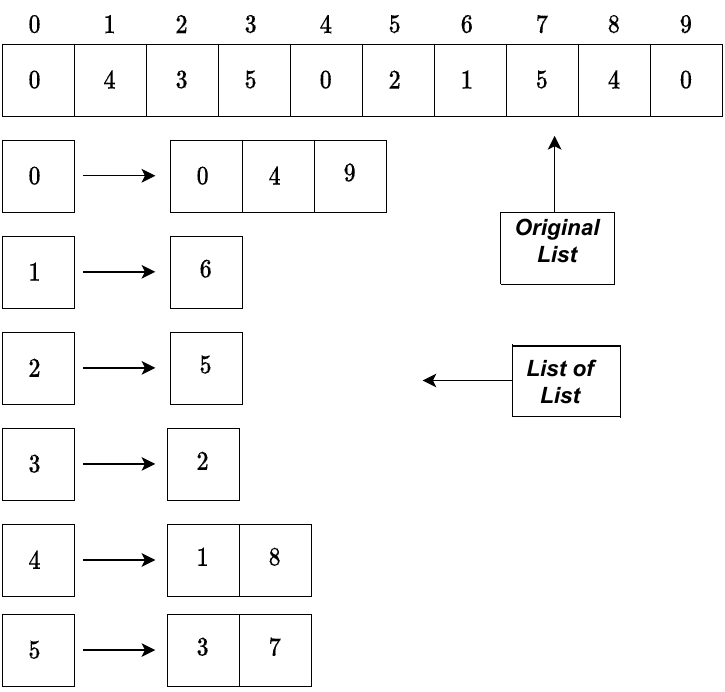} 
    \caption{List of List}
    \label{fig:listlist}
\end{figure}

The \texttt{owner} is a list with a size equal to the number of faces whose elements are the labels identifying the cell owner of the face. The elements in the list vary in the range $[0,N_{cell}]$, where $N_{cell}$ is the number of cells. Since each cell label owns different faces, the loop from Listing~\ref{lst:firstcycle} can be safely parallelised by iterating over the number of cells and for each cell iterating over the faces that this cell owns. One way to organise the owner list in this way is to reorder the elements in a list of lists, see Figure~\ref{fig:listlist}, in which the external list represents the number of cells and the internal lists represent the faces each cell owns.

\vspace{0.25cm}
\

\begin{figure}[h!]
    \centering
    \includegraphics[scale=0.45]{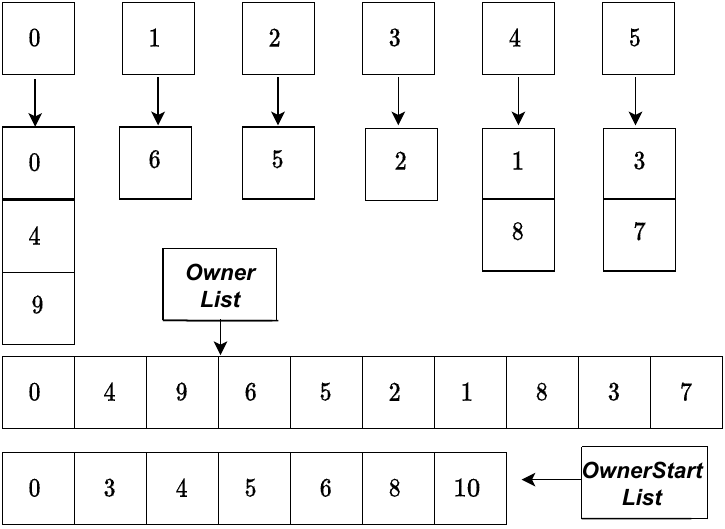} 
    \caption{Compressed format of a list of list}
    \label{fig:listlist2}
\end{figure}

Taking inspiration from RapidCFD\footnote{\url{https://github.com/Atizar/RapidCFD-dev}}, one way to build these lists is to create a first list with the elements contained in the internal list and another one that contains the starting point of the previous list between two successive cell elements, see Figure~\ref{fig:listlist2}. The challenge is to create these lists in parallel using modern C++ techniques. It is important to note that, in a general OpenFOAM simulation, these lists are used multiple times during the simulation, but unless the mesh geometry changes, they are only created once. 

\vspace{0.25cm}

\begin{figure}[h!]
    \centering
    \includegraphics[scale=0.45]{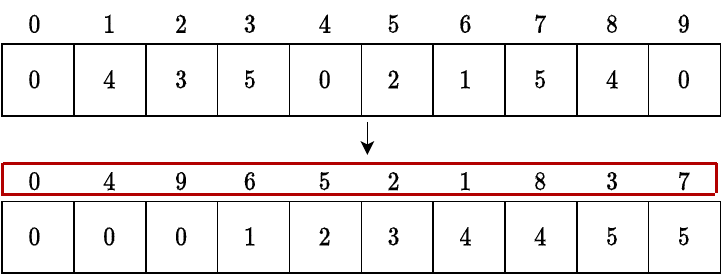} 
    \caption{Sorting owner list}
    \label{fig:listlist3}
\end{figure}

The standard sort algorithm can efficiently compute the first list in parallel. Specifically, two lists are needed: the original owner list and a second one, which is the output and is initialized as a list of indices from $0$ to the size of the owner list. The wanted output is obtained by sorting the first list and rearranging the second one accordingly, see Figure~\ref{fig:listlist3}.

\vspace{0.25cm}

\begin{figure}[h!]
    \centering
    \includegraphics[scale=0.40]{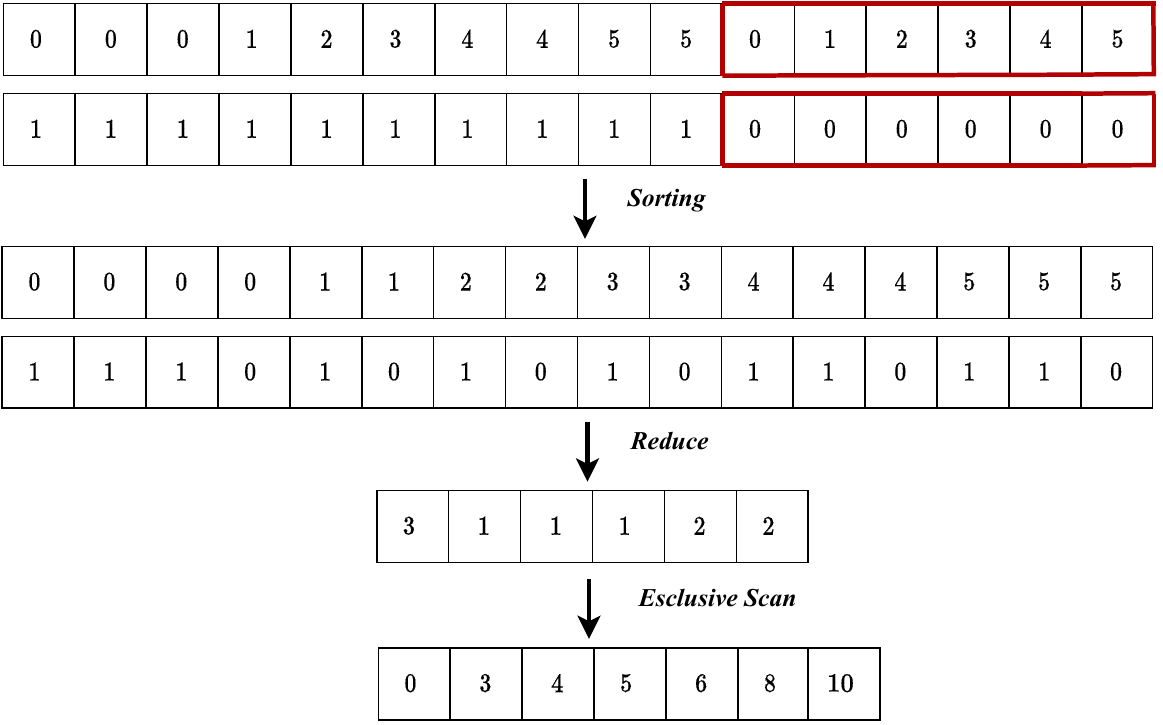} 
    \caption{Starting index of the previous list}
    \label{fig:listlist4}
\end{figure}

The second list is slightly more challenging to compute. Starting from the reordered owner list, another list is appended at the end. This list has indices from $0$ to the number of cells $N_{cell}$. Another list of size owner list size plus $N_{cell}$ is created, initializing its elements to zero. Then, all elements from $0$ to the owner size are set to $1$. The second step is sorting the lists as previously. The third step is to reduce both lists according to the first list while the elements of the second list are added. Finally an \texttt{exclusive\_scan} algorithm can be applied (Figure~\ref{fig:listlist4}).

\vspace{0.25cm}

These lists are helpful to parallelize the first gradient loop safely. To compute that loop in parallel, four lists are needed: the \texttt{ownerList\_}, the \texttt{ownerStart\_}, the \texttt{neighbourList\_} and the \texttt{neighbourStart\_}. We insert these lists as \texttt{std::unique\_ptr} in the class \texttt{lduAddressing} coherently with the lists defined in that class by the actual OpenFOAM version. Public functions \texttt{ownerList}, \texttt{ownerStart}, \texttt{neighbourList}, and \texttt{neighbourStart} have been defined in order to be able to construct these lists from other sections of the code, i.e. from the \texttt{gradf} function. 

With these new lists, the loop \ref{lst:firstcycle} is rewritten as:

\vspace{0.2cm}
\vbox{\lstset{style=openfoamStyle}
\begin{lstlisting}[language=C++,caption={First cycle gradient in parallel}, label={lst:firstcyclepar}]
const labelUList& owlist=mesh.lduAddr().ownerList();
const labelUList& owstart=mesh.lduAddr().ownerStart();
const labelUList& nelist=mesh.lduAddr().neighbourList();
const labelUList& nestart=mesh.lduAddr().neighbourStart();

auto iter=std::views::iota(0,igGrad.size());
std::for_each(std::execution::par,iter.begin(),iter.end(),
    [ol=owlist.cdata(),os=owstart.cdata(),nl=nelist.cdata(),ns=nestart.cdata(),sf=Sf.cdata(),is=issf.cdata(),ig=igGrad.data()](const label& facei){
        ig[facei]=Zero;
        for(int i=os[facei]; i<os[facei+1];++i){
            ig[facei]+= sf[ol[i]]*is[ol[i]];
        }
        for(int i=ns[facei]; i<ns[facei+1];++i){
            ig[facei]-= sf[nl[i]]*is[nl[i]];
        }
    });
\end{lstlisting}} 

The second loop in \texttt{gradf} takes into account the boundary of the domain; see Listing~\ref{lst:secondcycle}.

\vspace{0.2cm}
\vbox{\lstset{style=openfoamStyle}
\begin{lstlisting}[language=C++,caption={Second cycle gradient}, label={lst:secondcycle}]
forAll(mesh.boundary(), patchi)
{
    const labelUList& pFaceCells =mesh.boundary()[patchi].faceCells();
    const vectorField& pSf = mesh.Sf().boundaryField()[patchi];
    const fvsPatchField<Type>& pssf = ssf.boundaryField()[patchi];
    forAll(mesh.boundary()[patchi], facei)
    {
        igGrad[pFaceCells[facei]] += pSf[facei]*pssf[facei];
    }
}
\end{lstlisting}}

The most computationally intensive loop is the innermost one, so the goal is to parallelize it. However, as in the case of the first loop, running the innermost loop in parallel without any precautions would lead to a data race. To parallelize it, we use the same procedure as for the first loop using lists of lists. These new lists can be created once. To simplify data management we decided to calculate them on the fly.

\vspace{0.25cm}
\begin{figure}[h!]
    \centering
    \includegraphics[scale=0.45]{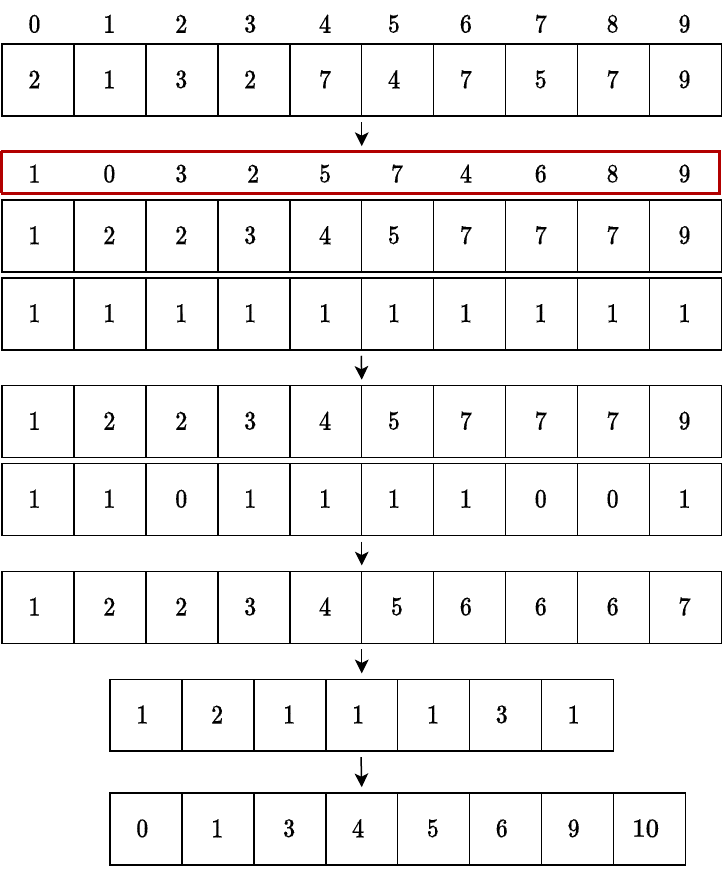} 
    \caption{Compressed lists for the boundaries}
    \label{fig:listlist5}
\end{figure}

Their implementation is in the \texttt{fvboundaryMesh.C} class. Similar to before, we created two lists of lists \texttt{facePatchIndex\_} and \texttt{facePatchStart\_} and the methods \texttt{facePatchIndex} and \texttt{facePatchStart} to construct them. The first list is obtained following the sorting procedure seen before. The other is obtained by first initializing the list to one, then iterating over the sorted list. When two adjacent elements are equal, the element in the list of ones corresponding to the index of the second element found to be equal is set to $0$. Then, an \texttt{inlusive\_scan} is applied. Iterating over this list, we save the counts of the repeated elements on another list, and finally we perform an \texttt{exclusive\_scan} on that list to obtain the expected result (Figure~\ref{fig:listlist5}). With the new lists, the innermost loop in Listing ~\ref{lst:secondcycle} can be parallelized as in Listing \ref{lst:secondparcycle}. This rewriting avoids race conditions and the need for strictly ordered iteration.

\vspace{0.2cm}
\vbox{\lstset{style=openfoamStyle}
\begin{lstlisting}[language=C++,  caption={Second cycle in parallel}, label={lst:secondparcycle}]
const auto& faceIndex=mesh.boundary().facePatchIndexPatch(patchi, mesh.boundary());
const auto& faceStart=mesh.boundary().facePatchStartPatch(patchi, mesh.boundary());
std::for_each(std::execution::par,std::views::iota(0).begin(),std::views::iota(faceStart.size()-1).begin(), 
   [ig=igGrad.data(),f=pFaceCells.cdata(),fir=pSf.cdata(),sec=pssf.cdata(),pstr=faceStart.cdata(),
        plst=faceIndex.cdata()](const label& facei){
            label id=f[plst[pstr[facei]]];                           
            for(int i=pstr[facei]; i<pstr[facei+1];++i){
                ig[id]+=fir[plst[i]]*sec[plst[i]];
            }
        });
}   
\end{lstlisting}}

Another intensive computation inside \texttt{gradf} is the field division as highlighted in the Listing \ref{lst:thirdcycle}. 

\vspace{0.2cm}
\vbox{\lstset{style=openfoamStyle}
\begin{lstlisting}[language=C++,caption={Field division }, label={lst:thirdcycle}]
igGrad /= mesh.V();
\end{lstlisting}}

The \texttt{TFOR\_ALL\_F\_OP\_F\_OP\_F} in \texttt{FieldM.H} has been modified as per Listing \ref{lst:thirdparcycle}.

\vspace{0.2cm}
\vbox{\lstset{style=openfoamStyle}
\begin{lstlisting}[language=C++,caption={Field division in parallel }, label={lst:thirdparcycle}]
#define TFOR_ALL_F_OP_F_OP_F(typeF1, f1, OP1, typeF2, f2, OP2, typeF3, f3) \
/* Check fields have same size */                       \
checkFields(f1, f2, f3, "f1 " #OP1 " f2 " #OP2 " f3");  \
/* Field access */                                      \
List_ACCESS(typeF1, f1, f1P);                           \
List_CONST_ACCESS(typeF2, f2, f2P);                     \
List_CONST_ACCESS(typeF3, f3, f3P);                     \
/* Loop: f1 OP1 f2 OP2 f3 */                            \
const label loopLen = (f1).size();                      \
                                                        \
std::for_each(std::execution::par,                      \
                std::views::iota(0, loopLen).begin(),   \
                std::views::iota(0, loopLen).end(),     \
                [=](const auto& i){                     \
                    (f1P[i]) OP1 (f2P[i]) OP2 (f3P[i]); \
                });                             
\end{lstlisting}}

\subsubsection{CorrectBoundaryConditions}

Listing~\ref{lst:bc_ori} shows the main cycle of 
\texttt{correctBoundary\-Conditions}.
The main compute involves the field operations, both calling the \texttt{BINARY\_OPERATOR} of the class \texttt{FieldFunctions.C}, which is rewritten in modern ISO C++ using PSTL to execute offload automatically on GPUs.


\vspace{0.2cm}
\vbox{\lstset{style=openfoamStyle}
\begin{lstlisting}[language=C++,caption={ CorrectBoundaryConditions }, label={lst:bc_ori}]
forAll(vsf.boundaryField(), patchi)
{
    if (!vsf.boundaryField()[patchi].coupled())
    {
        const vectorField n
        (
            vsf.mesh().Sf().boundaryField()[patchi]
          / vsf.mesh().magSf().boundaryField()[patchi]
        );
        gGradbf[patchi] += n *
        (
            vsf.boundaryField()[patchi].snGrad()
          - (n & gGradbf[patchi])
        );
    }
}
\end{lstlisting}}
\section{Evaluation}

\begin{figure*}[htbp]
    \centering
    \begin{subfigure}[b]{0.24\textwidth}
        \centering
        \includegraphics[width=\linewidth]{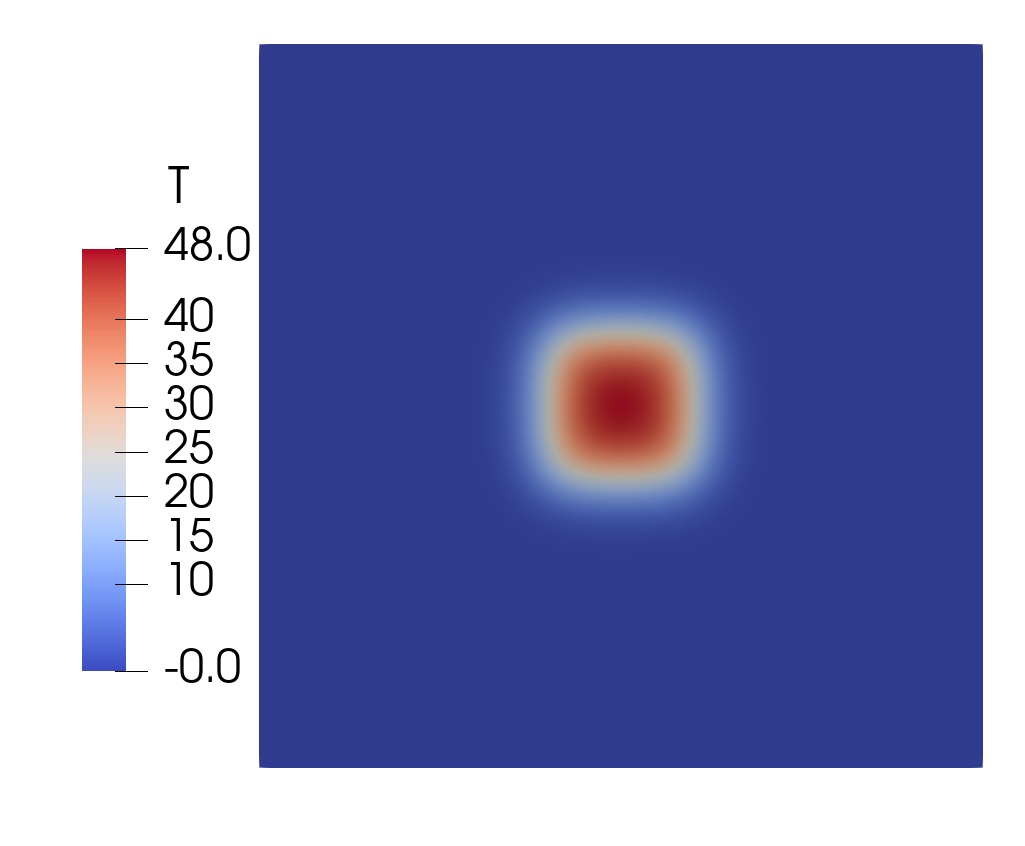}
        \caption{Temperature Frame 2s}
    \end{subfigure}\hfill
    \begin{subfigure}[b]{0.24\textwidth}
        \centering
        \includegraphics[width=\linewidth]{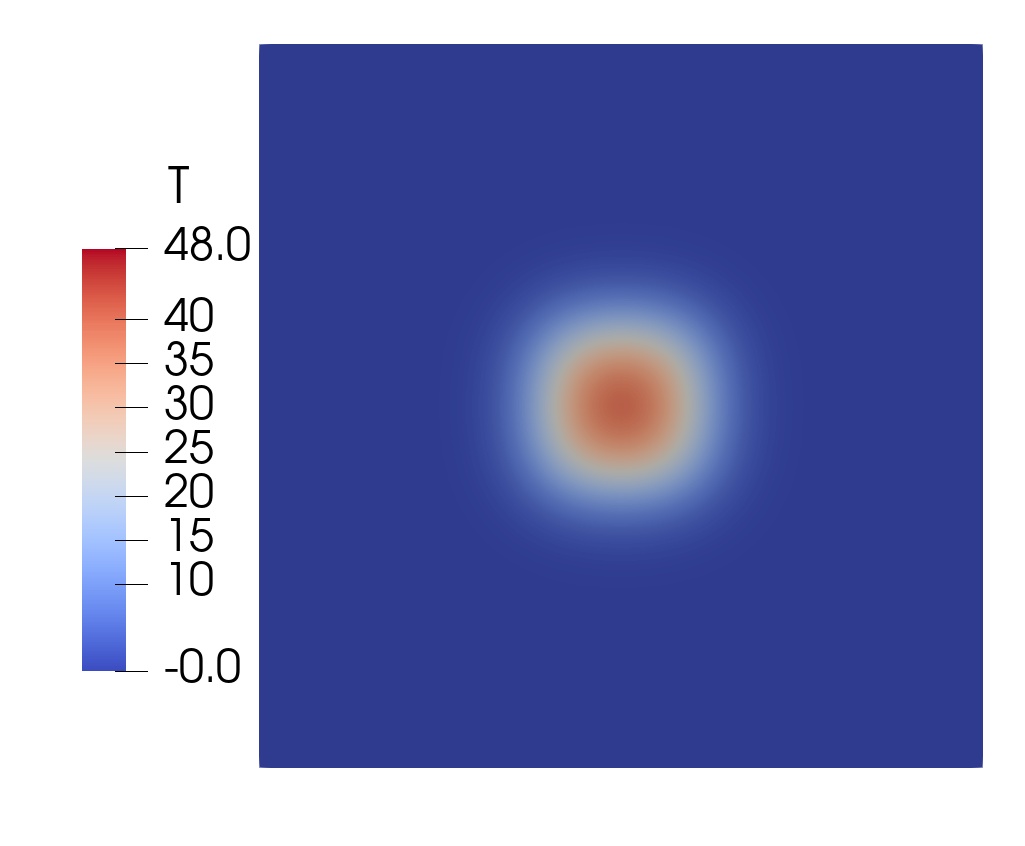}
        \caption{Temperature Frame 4s}
    \end{subfigure}\hfill
    \begin{subfigure}[b]{0.24\textwidth}
        \centering
        \includegraphics[width=\linewidth]{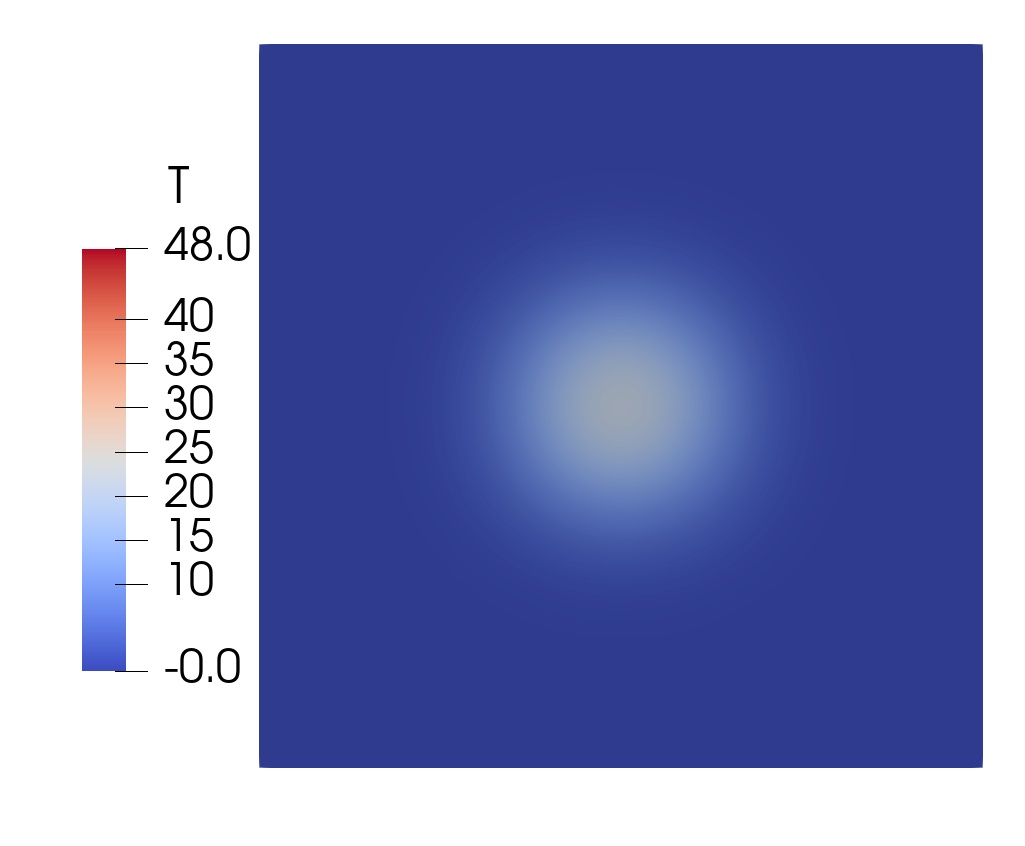}
        \caption{Temperature Frame 10s}
    \end{subfigure}\hfill
    \begin{subfigure}[b]{0.24\textwidth}
        \centering
        \includegraphics[width=\linewidth]{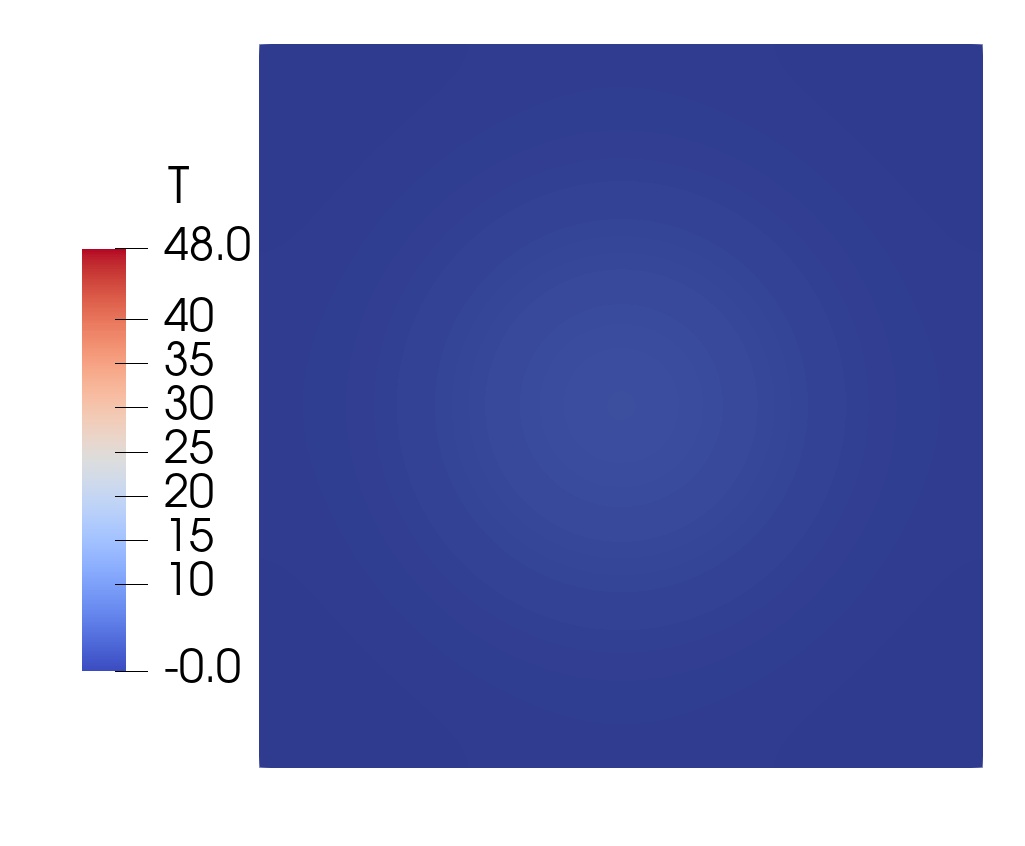}
        \caption{Temperature Frame 50s}
    \end{subfigure}
    \caption{Simulation used to test our Proof-Of-Concept}
    \label{fig:sim}
\end{figure*}

We use the \laplacianfoam solver (see Section ~\ref{laplacianFOAM}) for our proof of concept.
To test our implementation, starting from the same 3D block domain, we create four meshes of increasing sizes. Table~\ref{tab:meshes} shows the specifics of these meshes obtained using the OpenFOAM \texttt{checkMesh} application. Figure~\ref{fig:sim} shows the solution of the simulation on the \meshs at different time steps obtained from our parallel implementation. We verify the accuracy of the numerical results of our proof of concept by comparing them with the results obtained using the original OpenFOAM-24.12 release. 

\begin{table}[htbp]
    \centering
    \caption{Mesh dimensions.}
    \label{tab:meshes}
    \begin{small}
    \begin{tabular}{|c|c|c|c|c|}
        \hline
        \textbf{Property} & \textbf{\meshs} & \textbf{\meshm} & \textbf{\meshl} & \textbf{\meshxl} \\
        \hline
        Dimension             & $(100)^3$     & $(200)^3$     & $(300)^3$     & $(400)^3$     \\
        nCell                 & $1M$       & $8M$       & $27M$      & $64M$      \\
        Points                & $\sim1M$       & $\sim8M$       & $\sim27M$      & $\sim64M$      \\
        Faces                 & $\sim3M$       & $\sim24M$      & $\sim81M$      & $\sim192M$     \\
        Internal faces        & $\sim2M$       & $\sim23M$      & $\sim80M$      & $\sim191M$     \\
        \hline
    \end{tabular}
    \end{small}
\end{table}

In our simulations, we did not save the solution at each time step since it is a parameter that strictly depends on the user's purpose and analysing its impact on simulation is not the scope of this work. Instead, we consider three-time intervals: the time taken to execute the assembly phase, the time taken for the solver, and the total execution time, which also takes care of the time to read the initial data. We simulate the first 100 seconds, using a time step of 0.2 seconds. The selected solver is the Conjugate Gradient method (\texttt{PCG}) with a \texttt{diagonal} precondition. Each simulation is executed five times to increase the results' statistics.

Table~\ref{tab:platform} lists the architectures used. This work considers only NVIDIA GPUs; specifically, we use NVIDIA V100 (\cascadelake), A100 (\epito), H100 (\twophases) and H200 (\gracehopper). These systems are hosted by the HPC4AI~\cite{HPC4AI} - High Performance Computing for Artificial Intelligence - infrastructure hosted and operated by University of Turin (Italy).

We compile both versions of OpenFOAM 24.12, official release and our proof of concept, with the \texttt{nvc++-24.3} compiler and GNU version \texttt{12.2.0} using the default optimization flag \texttt{-O3}. Libraries \texttt{libopenFOAM.so} and \texttt{libfiniteVolume.so} and \texttt{libfvOptions.so} are build by dding the flags \texttt{-cuda -stdpar -gpu=\$\{COMPUTE\_CAPABILITY\},managed}. These flags turn on \texttt{nvc++} PSTL support and compile for GPU offload. The \texttt{\$\{COMPUTE\_CAPABILITY\}} varies based on the GPU architecture. 

\begin{table*}
\caption{Systems specifications}
\scalebox{0.75}{
\begin{tabular}{|l|c|c|c|c|}
\hline 
\textbf{System} & \textbf{\cascadelake} & \textbf{\epito} & \textbf{ \gracehopper} & \textbf{\twophases} \\
 \hline
 CPU Manufacturer & Intel(R) & Ampere Computing  & NVIDIA & Intel(R) \\
 CPU Model &  Xeon Gold 6230 & Q80-30  & Grace & Xeon Platinum 8458P\\
 CPU Codename & Cascade Lake & Altra & GH200 & Sapphire Rapids \\
 CPU uArch & Skylake & ARM Neoverse N1 & ARM Neoverse-V2 & Golden Cove\\
 CPU cores per socket & 20 & 80 & 72 & 44\\
 CPU Sockets & 2 & 1 & 1 & 2\\
 CPU cores per node & 40 & 80 & 72 & 88\\
 \hline
 GPU Architecture & Volta & Ampere & Hopper & Hopper \\
 GPU Model & V100 & A100 & H100 & H100 \\
 GPU Compute Capability & 70 & 80 & 90 & 90 \\
 GPU Memory [GB] & 32 & 40 & 96 & 82 \\
 GPU-CPU Connectivity & PCIe gen3 & PCIe gen4 & NVLink-C2C & PCIe gen4 \\
 GPU-GPU Connectivity & n/a & n/a & n/a & NVLink \\
 GPU per node & 1 & 2 & 1 & 4\\
 GPU Memory per node [GB] & 32 & 80 & 96 & 328 \\
 \hline
\end{tabular}
}
\label{tab:platform}
\end{table*}


\subsection{Performance results on single GPU}

\begin{table*}[ht]
\small
\caption{Performance comparison across GPU architectures and configurations for different mesh sizes.}
\resizebox{\textwidth}{!}{
\begin{tabular}{|c|c|ccc|ccc|ccc|ccc|}
\hline
\textbf{Mesh} & \textbf{Metric} & \multicolumn{3}{c|}{\textbf{\cascadelake}} & \multicolumn{3}{c|}{\textbf{\epito}} & \multicolumn{3}{c|} {\textbf{\gracehopper}} & \multicolumn{3}{c|}{\textbf{\twophases}} \\
& & 40 MPI & GPU & GPU+A & 40 MPI & GPU & GPU+A & 72 MPI & GPU & GPU+ALL & 88 MPI & GPU & GPU+A \\
\hline

\multirow{3}{*}{\meshs}
& Assembly (s)  & 3.18\scalebox{.6}{$\pm0.11$} & 1.19\scalebox{.6}{$\pm0.06$} & 1.13\scalebox{.6}{$\pm0.04$} 
                 & 5.40\scalebox{.6}{$\pm0.46$} & 0.89\scalebox{.6}{$\pm0.09$} & 0.95\scalebox{.6}{$\pm0.06$} 
                 & 1.44\scalebox{.6}{$\pm0.08$} & 0.53\scalebox{.6}{$\pm0.11$} & 0.56\scalebox{.6}{$\pm0.08$} 
                 & 1.19\scalebox{.6}{$\pm0.11$} & 0.60\scalebox{.6}{$\pm0.08$} & 0.53\scalebox{.6}{$\pm0.060$} \\
& Solver (s)    & 3.64\scalebox{.6}{$\pm0.10$} & 3.44\scalebox{.6}{$\pm0.09$} & 3.60\scalebox{.6}{$\pm0.09$}
                 & 5.34\scalebox{.6}{$\pm0.38$} & 2.87\scalebox{.6}{$\pm0.11$} & 2.77\scalebox{.6}{$\pm0.07$}
                 & 1.64\scalebox{.6}{$\pm0.11$} & 1.98\scalebox{.6}{$\pm0.11$} & 1.93\scalebox{.6}{$\pm0.11$}
                 & 0.85\scalebox{.6}{$\pm0.08$} & 2.00\scalebox{.6}{$\pm0.09$} & 2.04\scalebox{.6}{$\pm0.05$} \\
& Execution (s) & 7.20\scalebox{.6}{$\pm0.02$} & 12.42\scalebox{.6}{$\pm0.29$} & 12.51\scalebox{.6}{$\pm0.16$}
                 & 11.05\scalebox{.6}{$\pm0.11$} & 13.82\scalebox{.6}{$\pm0.21$} & 13.74\scalebox{.6}{$\pm0.08$}
                 & 3.37\scalebox{.6}{$\pm0.05$} & 6.69\scalebox{.6}{$\pm0.02$} & 6.78\scalebox{.6}{$\pm0.02$}
                 & 2.59\scalebox{.6}{$\pm0.05$} & 8.10\scalebox{.6}{$\pm0.04$} & 8.14\scalebox{.6}{$\pm0.02$} \\

\hline
\multirow{3}{*}{\meshm}
& Assembly (s)  & 35.02\scalebox{.6}{$\pm0.67$} & 5.82\scalebox{.6}{$\pm0.13$} & 5.93\scalebox{.6}{$\pm0.11$}
                 & 
                 40.38\scalebox{.6}{$\pm1.10$} & 4.53\scalebox{.6}{$\pm0.09$} & 4.44\scalebox{.6}{$\pm0.12$}
                 & 17.50\scalebox{.6}{$\pm0.09$} & 1.95\scalebox{.6}{$\pm0.05$} & 2.04\scalebox{.6}{$\pm0.06$}
                 & 19.25\scalebox{.6}{$\pm0.46$} & 2.30\scalebox{.6}{$\pm0.05$} & 2.20\scalebox{.6}{$\pm0.08$} \\
& Solver (s)    & 95.03\scalebox{.6}{$\pm0.72$} & 29.95\scalebox{.6}{$\pm0.14$} & 29.76\scalebox{.6}{$\pm0.09$}
                 & 105.55\scalebox{.6}{$\pm1.49$} & 25.11\scalebox{.6}{$\pm0.11$} & 25.19\scalebox{.6}{$\pm0.12$}
                 & 47.92\scalebox{.6}{$\pm0.20$} & 12.59\scalebox{.6}{$\pm0.04$} & 12.53\scalebox{.6}{$\pm0.06$}
                 & 33.21\scalebox{.6}{$\pm0.66$} & 14.7\scalebox{.6}{$\pm0.09$} & 14.75\scalebox{.6}{$\pm0.07$} \\
& Execution (s) & 133.16\scalebox{.6}{$\pm0.35$} & 97.35\scalebox{.6}{$\pm0.41$} & 95.14\scalebox{.6}{$\pm0.17$}
                 & 153.57\scalebox{.6}{$\pm0.99$} & 111.83\scalebox{.6}{$\pm0.97$} & 107.43\scalebox{.6}{$\pm0.68$}
                 & 66.81\scalebox{.6}{$\pm0.13$} & 49.78\scalebox{.6}{$\pm0.12$} & 50.83\scalebox{.6}{$\pm0.26$}
                 & 52.85\scalebox{.6}{$\pm0.62$} & 63.10\scalebox{.6}{$\pm0.24$} & 63.2\scalebox{.6}{$\pm0.17$} \\

\hline
\multirow{3}{*}{\meshl}
& Assembly (s)  & 126.88\scalebox{.6}{$\pm0.63$} & 153.25\scalebox{.6}{$\pm0.19$} & 189.83\scalebox{.6}{$\pm1.92$}
                 & 149.82\scalebox{.6}{$\pm0.26$} & 236.70\scalebox{.6}{$\pm0.38$} & 14.20\scalebox{.6}{$\pm0.07$}
                 & 63.59\scalebox{.6}{$\pm0.30$} & 108.09\scalebox{.6}{$\pm0.43$} & 5.90\scalebox{.6}{$\pm0.50$}
                 & 63.60\scalebox{.6}{$\pm2.13$} & 142.50\scalebox{.6}{$\pm1.42$} & 6.94\scalebox{.6}{$\pm0.12$} \\
& Solver (s)    & 464.12\scalebox{.6}{$\pm0.98$} & 187.51\scalebox{.6}{$\pm0.24$} & 217.26\scalebox{.6}{$\pm1.47$}
                 & 516.98\scalebox{.6}{$\pm0.19$} & 103.12\scalebox{.6}{$\pm0.06$} & 104.13\scalebox{.6}{$\pm0.04$}
                 & 291.38\scalebox{.6}{$\pm1.35$} & 48.81\scalebox{.6}{$\pm0.12$} & 48.76\scalebox{.6}{$\pm0.43$}
                 & 245.61\scalebox{.6}{$\pm2.69$} & 59.50\scalebox{.6}{$\pm0.03$} & 59.75\scalebox{.6}{$\pm0.07$} \\
& Execution (s) & 597.91\scalebox{.6}{$\pm1.33$} & 549.37\scalebox{.6}{$\pm0.46$} & 622.28\scalebox{.6}{$\pm6.17$}
                 & 678.41\scalebox{.6}{$\pm0.47$} & 635.43\scalebox{.6}{$\pm1.69$} & 396.27\scalebox{.6}{$\pm4.84$}
                 & 359.46\scalebox{.6}{$\pm1.39$} & 280.96\scalebox{.6}{$\pm1.93$} & 183.95\scalebox{.6}{$\pm0.73$}
                 & 317.30\scalebox{.6}{$\pm0.89$} & 360.80\scalebox{.6}{$\pm1.42$} & 224.59\scalebox{.6}{$\pm0.97$} \\

\hline
\multirow{3}{*}{\meshxl}
& Assembly (s)  & \multicolumn{3}{c|}{--} 
                 & \multicolumn{3}{c|}{--} 
                 &  148.29\scalebox{.6}{$\pm1.01$} & 250.06\scalebox{.6}{$\pm0.92$} & 13.31\scalebox{.6}{$\pm0.08$} 
                 & 155.35\scalebox{.6}{$\pm2.53$} & 333.80\scalebox{.6}{$\pm0.21$} & 15.65\scalebox{.6}{$\pm0.17$} \\
& Solver (s)    & \multicolumn{3}{c|}{--}  
                 & \multicolumn{3}{c|}{--} 
                 & 915.66\scalebox{.6}{$\pm3.87$} & 163.81\scalebox{.6}{$\pm0.10$} & 144.92\scalebox{.6}{$\pm0.11$} 
                 & 791.61\scalebox{.6}{$\pm2.78$} & 204.00\scalebox{.6}{$\pm0.19$} & 177.17\scalebox{.6}{$\pm0.22$} \\
& Execution (s) & \multicolumn{3}{c|}{--} 
                 & \multicolumn{3}{c|}{--} 
                 &  1073.93\scalebox{.6}{$\pm3.65$} & 699.85\scalebox{.6}{$\pm1.74$} & 458.93\scalebox{.6}{$\pm2.67$} 
                 & 957.15\scalebox{.6}{$\pm0.42$} & 907.20\scalebox{.6}{$\pm0.87$} & 560.29\scalebox{.6}{$\pm1.73$} \\
\hline
\end{tabular}
}
\label{tab:single_gpu_perf}
\end{table*}

Table~\ref{tab:single_gpu_perf} shows the performance of the \laplacianfoam application for different architectures and mesh sizes. Specifically, we compare the best MPI results (first column for each architecture). Multiple tests running only MPI have been performed; we chose to report the best MPI count, corresponding to all cases except one (\epito) of all available cores on the system. 

The acceleration of the assembly phase in the case of \meshs varies from a minimum of $1.98$x to a maximum of $2.82$x across the \cascadelake, \gracehopper, and \twophases systems. An exception comes from \epito, where the speed-up is $5.6$x. While a deeper investigation is left for future work, our main hypothesis is that the cores of the Arm Neoverse N1 CPU are less performant with respect to the cores available on the other platforms. This is motivated by the fact that the GPU performance behaves as expected, with the slowest execution time achieved of $1.19$s on V100 (the oldest GPU considered in this work) compared to $0.89$s on A100, $0.60$s on H100, and $0.53$s on H200.

In the case of \meshm, the speed-up resulting from the assembly phase is generally higher compared to the previous case. Specifically, the\gracehopper achieves a speed-up of $8.96$x when utilizing a single GPU compared to a 72-MPI configuration. Similarly, both A100 and V100 attain speed-ups of $8.91$x and $6.02$x, respectively, relative to a 40-MPI baseline. On the \twophases system, the assembly phase demonstrates an $8.44$x performance improvement when executed on a single GPU versus 88 MPI.

In the case of \meshl, the assembly performance drops to $0.45$x and $0.59$x in the \twophases and \gracehopper systems, respectively. The same trend is seen in the other architectures, with the performance of $0.56$x and $0.83$x, comparing 40 MPI versus 1 GPU on the A100 and V100, respectively. This behaviour is due to the memory allocation and management done by nvc++ and the CUDA Driver. Generally, it uses a memory allocator of a predefined size by default. The memory allocated and managed by this allocator is released once at the end of the simulation. However, when the assembly field sizes are too big to fit in this memory, the deallocation happens at the end of every assembly phase, degrading the performance. This is easily seen using the Nsight System profiler which allows to visualize automatic memory page movements done by the driver runtime. Varying the memory pool allocator's behaviour and pre-defined size (by setting appropriate environmental variables \texttt{NVCOMPILER\_ACC\_POOL\_ALLOC},
\texttt{NVCOMPILER\_ACC\_POOL\_SIZE}, and
\texttt{NVCOMPILER\_ACC\_POOL\_THRESHOLD}) mitigates this effect on all the architectures except for \cascadelake, where \meshl is still too big. Although interesting, this work is not focused on deeply investigating the best allocator size setting and leave this analysis for future work. 

In Table~\ref{tab:single_gpu_perf}, the column {"GPU+A"} represents the simulation results obtained by setting the allocator sizes. We set \texttt{NVCOMPILER\_ACC\_POOL\_THRESHOLD=100} to let the memory pool allocator occupy the entire device memory. As stated before, on Cascadelake, even setting \texttt{NVCOMPILER\_ACC\_POOL\_ALLOC} and \texttt{NVCOMPILER\_ACC\_POOL\_SIZE} to 30 GB (close to maximum capacity for that GPU), the mesh is too big to see performance improvement. On \epito, setting them to 40 GB, we achieved a speed-up of $10.55$x on one GPU versus 40 MPI. On \gracehopper and \twophases, we achieved speedups of $10.78$x and $9.16$x compared to 72 and 88 MPI by setting the variables to 10 GB.

\text{\meshxl}’s results show a similar patter of \meshl on \gracehopper and \twophases. Without setting the memory pool allocator explicitly, the assembly step performance drops to $0.59$x and $0.47$x compared to 72 and 88 MPI. Setting the memory pool allocator size unlock visible performance gains $11.14$x and $9.93$x, respectively.

\begin{figure*}[htbp]
    \centering
    \includegraphics[width=\linewidth]{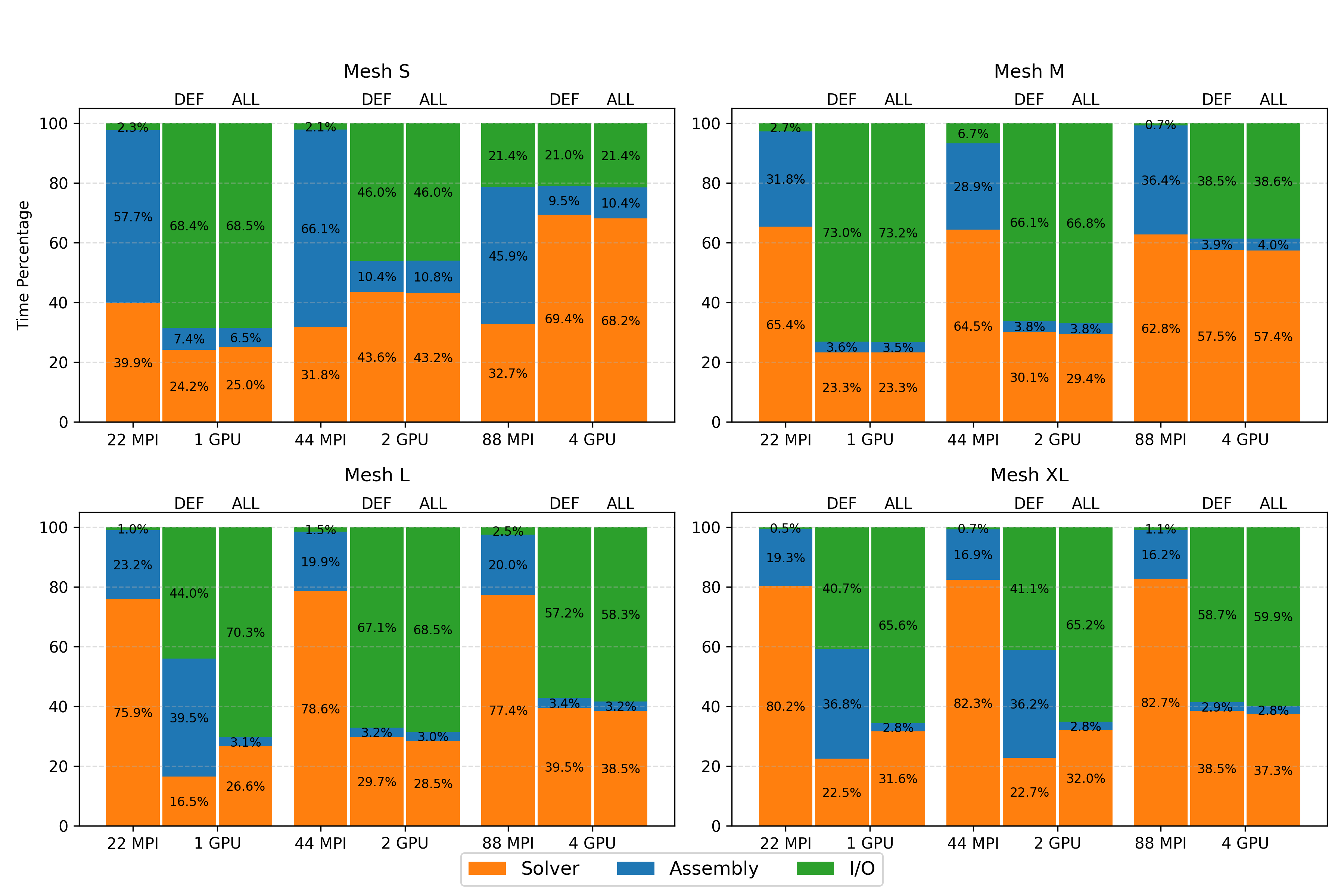}
    \caption{Multi-GPUs simulations on \texttt{TwoPhases}.  \texttt{DEF} and \texttt{ALL} labels indicate simulations executed with the default and 10 GB allocator sizes.}
    \label{fig:multigpus_percentage}
\end{figure*}

Focusing on the \laplacianfoam application end-to-end behavior, the initialization time to read and create fields (labeled as "I/O") represents a significant bottleneck when running simulations on GPUs. Figure \ref{fig:multigpus_percentage} shows the impact of I/O on \twophases: when GPUs are not used, I/O represents less than 5\% of the entire simulation time. When \texttt{Solver} and \texttt{Assembly} are offloaded to the GPUs, I/O becomes the dominant non-accelerated portion of the code, dramatically reducing the end-to-end speedup of the application. This is not a surprise, as Amdhal's law dictate.


\subsection{Performance results running multple GPUs}

\begin{figure*}[htbp]
    \centering
    \includegraphics[width=\linewidth]{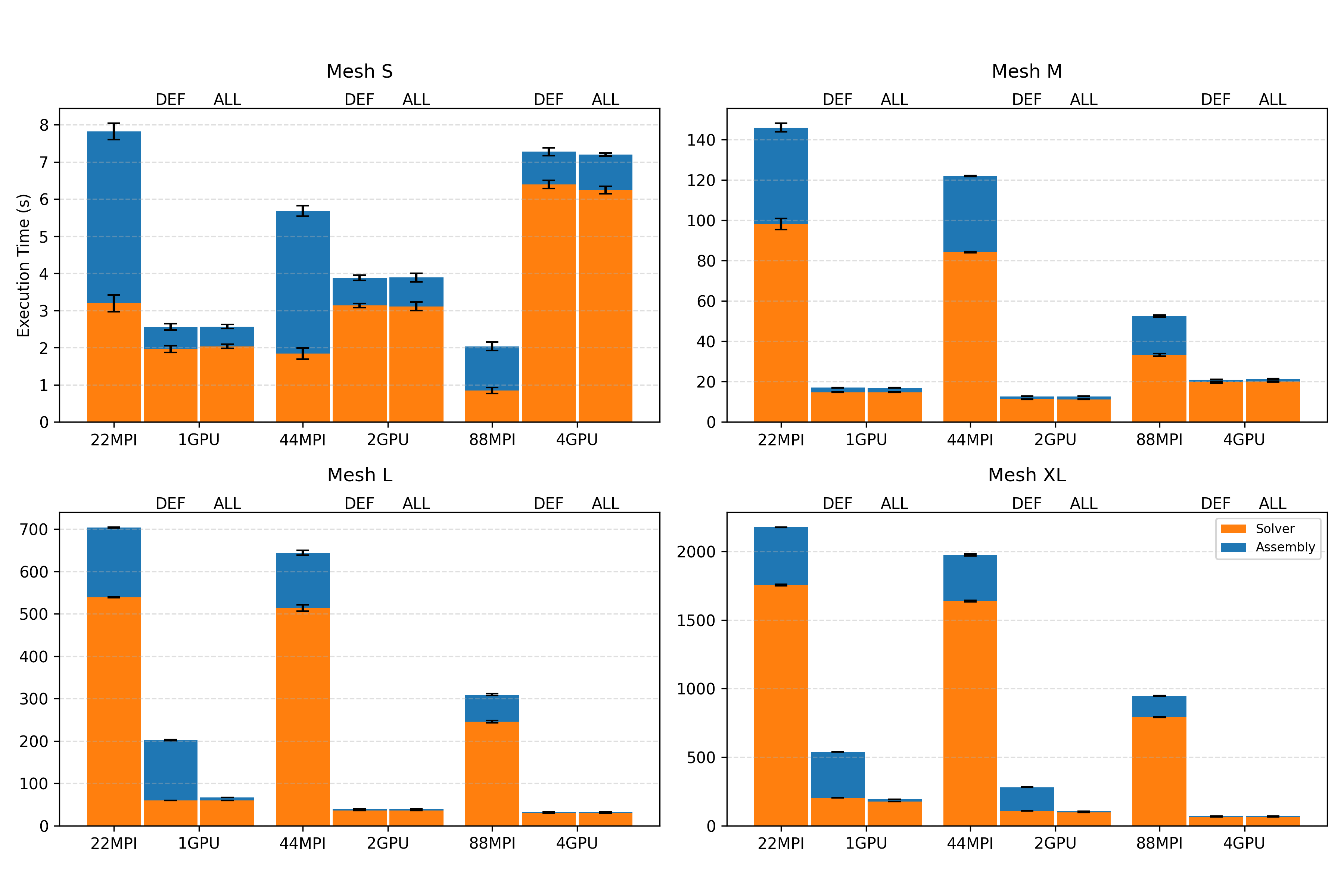}
    \caption{Multi-GPUs simulations on \texttt{TwoPhases}. \texttt{DEF} and \texttt{ALL} labels indicate simulations executed with the default and 10 GB allocator sizes. I/O time is not considered.}
    \label{fig:multigpus}
\end{figure*}

Figure~\ref{fig:multigpus} shows a performance comparison of our application executed using 1, 2, and 4 H100 GPUs of the \twophases machine (1 MPI process for each GPU). As before, \meshs is too small to profit from GPU offloading. In this case, we see that the overall performance slightly increases when executing on 2 GPUs and decreases on 4 GPUs. This improvement comes mainly from the I/O time (green in the figure), which increases from $5.55$s (1 GPU) to $3.33$s (2 GPU). In contrast, the solver time increases from $1.96$s to $3.14$s. This time gets even worse when running the simulation on 4 GPUs. Since the solver is the phase where MPI communication is dominant, the overhead they produce is greater than the gain from offloading to multiple GPUs. A similar trend is observed for all simulations with \meshm: in this case on 2 GPUs, the solver time decreases from $14.73$s (using 1 GPU) to $11.3$s, but increases to $19.66$s on 4 GPUs.

The performance of the assembly and solvers of \meshl and \meshxl shows a different behavior. Considering \meshl, there is an overall speed-up of $3$x from 1 to 2 GPUs and of $1.59$x from 2 to 4 GPUs, respectively. The first speed-up mainly derives from the assembly time, which drops from $142.53$s on 1 GPU to $3.8$s on 2 GPUs. This performance increment is consistent with the assembly results obtained by comparing the execution on 1 GPU using the default allocator and setting it to 10 GB. As highlighted in the previous section, \meshl and \meshxl are too large to manage their allocation with the default allocator size; for this reason, the NVIDIA driver has to swap and free the memory at every iteration of the \texttt{SIMPLE} loop which degrades performance. Increasing the size of the allocator helps mitigate this problem. The speed-up obtained by setting the allocator to 10 GB in the assembly phase of \meshl is $20.5$x. In the case of 2 GPUs, the overall mesh size is divided between the GPUs and can be efficiently managed by the default allocator size, producing a speedup of $37.5$x. Passing to 4 GPUs, the assembly time further improves from $3.8$s to $2.53$s. Considering the performance obtained by setting the allocator size to 10 GB while varying the number of GPUs, we notice that the speed-up is lower than in the previous case. Specifically, it is $1.79$x and $1.61$x from 1 to 2 and 2 to 4 GPUs, respectively. Again, this is because setting the allocator size to 10 GB avoids data swap and memory freeing when the simulation is executed on 1 GPU. 

Similar results are observed on \meshxl. In particular, setting the allocator size to 10 GB significantly improves the performance of the assembly phase. Unlike the previous case, memory management cannot be handled efficiently by the default allocator size, even when using 2 GPUs. The assembly time decreases from $333.8$s (with 1 GPU) to $172.6$s (with 2 GPUs). With 4 GPUs, the assembly time drops further to $4.82$s, indicating that the memory managed by each GPU is efficiently handled with the default allocator size. When setting the allocator size to 10 GB, the observed speed-ups are $1.86$x from 1 to 2 GPUs and $1.76$x from 2 to 4 GPUs. The solver time scales similarly to the \meshl case. The overall execution time speed-up is $1.9$x and $2.85$x, considering the default allocator size and GPUs from 1 to 2 and 2 to 4, respectively. Setting the allocator size to 10 GB becomes $1.85$x and $1.74$x, respectively. 
\section{Conclusions and outlooks}

A complex application like OpenFOAM poses challenges and constraints in which porting strategies can be adopted and maintained by the developer community. This goal of this work is to demonstrate that ISO C++ is a valid approach to achieve multi-core execution and GPU acceleration while maintaining forward-looking portability.

We developed and analyzed a proof-of-concept application, \laplacianfoam, to demonstrate that the adoption of standard C++ parallelization techniques in the OpenFOAM framework is feasible and performance improvements can be achieved, with reasonable and targeted tweaks to the underlying algorithm, data structures, and execution flow.

The overall results show that when meshes that are too small, the gain from GPU execution is outweighed primarily by the overhead generated by the automatic manage memory management. GPU performance of our application is also strongly affected by the chosen memory pool allocator settings, varying from positive gains to even leading to slowdowns if the data size does not match. The worst case observed is on the \twophases machine using \meshs and running 4 MPI with 4 GPUs versus 88 MPI (full node) reports a speed-up of $0.28$x. A speed-up of $5.57$x is achieved on the same architectures using \meshxl and running 4 MPI with 4 GPUs versus 88 MPI (fully populated node).


The choice of maximum size of the memory pool allocator strongly influences the performance of the assembly phase. The best speed-up obtained is $11.14x$ on the \gracehopper system and \meshxl, while the minimum is $2.25x$ on \twophases system and \meshxl.

In this work, we did not explore GPU from different vendors, like AMD or Intel. In the future, we plan to explore the compiler frameworks AdaptiveCpp~\cite{AdaptiveCpp} and Roc-stdpar~\cite{rocstdpar} to truly test portability GPU technologies. The strong focus on NVIDIA GPU was motivated by hardware availability and compiler maturity.

We acknowledge the narrow scope of the end-to-end application used, porting other operators and classes to GPU using a similar approach will expand the number of applications that can benefit from the GPU offloading. OpenFOAM core developers are better equipped to carry on the software engineering work required to widely adopt in the entire codebase more modern ISO C++ (including regression testing and validation). From a performance standpoint, the naive port of the Preconditioned Conjugate Solver (PCG) allows to run the \laplacianfoam application end-to-end. A better alternative would be the \texttt{AmgX} library developed by NVIDIA via the \texttt{petsc4FOAM} plug-in.

This work compellingly demonstrates that modernizing complex C++ codes to efficiently target accelerator architectures is not only feasible, but can be achieved with clear success, even under carefully described constraints and a deliberately narrow focus. Our approach delivered positive results, validating our thesis and providing a robust proof-of-concept. The outcomes show that performance improvements can be realized without sacrificing code maintainability or developer productivity, a crucial consideration for long-term sustainability in scientific computing.

Using modern C++ features and the Parallel Standard Template Library, this methodology enables transparent performance enhancements. This means that as the underlying implementation evolves, further gains can be realized automatically, freeing developers to focus on advancing numerical methods and unlocking new scientific capabilities, rather than on low-level optimization details.

Importantly, while the present study focused on OpenFOAM, the approach is not limited to this codebase. Thus, this work serves as a one more successful demonstrator of the broader applicability of modernizing legacy C++ codes to take advantage of increasingly common heterogeneous computing environments.
\section{Acknowledgments}
This work has been supported by the Spoke 1 ``FutureHPC \& BigData'' of the ICSC--Centro Nazionale di Ricerca in High-Performance Computing, Big Data and Quantum Computing-and hosting entity, funded by European \text{}tit{Union—Next GenerationEU}; by the DYMAN project funded by the European Union - European Innovation Council under G.A. n. 101161930;

Giovanni Stabile was funded by the European Union (ERC, DANTE, GA-101115741). Views and opinions expressed are, however, those of the author(s) only and do not necessarily reflect those of the European Union or the European Research Council Executive Agency. Neither the European Union nor the granting authority can be held responsible for them.

All experiments in this work have been performed with an officially released version of the OpenFOAM code (v2412); we acknowledge its developers and contributors. The authors thank the University of Turin and HPC4AI for access to the benchmarking systems.
\section*{Data availability statement}
Data is available upon reasonable request.
\section*{Declaration of generative AI and AI-assisted technologies in the writing process.}
During the preparation of this work, the authors used ChatGPT and WriteFull only to improve the readability and language of the manuscript. After using this tool/service, the authors reviewed and edited the content as needed and took full responsibility for the content of the published article.

\newpage
\bibliographystyle{cas-model2-names}

\end{document}